\documentclass[pra, twocolumn, superscriptaddress]{revtex4}
\usepackage{amssymb}
\usepackage{amsmath}
\usepackage{graphicx}
\usepackage{dcolumn}
\usepackage{color}
\usepackage{bm}
\usepackage{subfigure}
\usepackage{amsfonts}

\begin{document}
\preprint{APS/123-QED}
\title{Investigating bound entangled two-qutrit  states via the best separable approximation}
\author{A.\ Gabdulin }
\affiliation{Department of Physics, School of Science and Humanities, Nazarbayev University, 53 Qabanbay Batyr Avenue, Nur-Sultan 010000, Kazakhstan}
\author{ A.\ Mandilara}
\affiliation{Department of Physics, School of Science and Humanities, Nazarbayev University, 53 Qabanbay Batyr Avenue, Nur-Sultan 010000, Kazakhstan}
\begin{abstract} 
We use the linear programming algorithm introduced by Akulin \textit{et al.} [V. M. Akulin, G. A. Kabatiansky,
and A. Mandilara, Phys. Rev. A 92, 042322 (2015)] to perform  best separable approximation  on  two-qutrit random density matrices. We combine the numerical results with theoretical methods in order to generate random  representative families of positive partial transposed bound entangled (BE) states and analyze their properties. Our  results are disclosing that for the two-qutrit system the  BE states have negligible volume and that these form tiny `islands' sporadically distributed  over the surface of the polytope of separable states. 
The detected families of  BE states are found to be located under  a layer of pseudo one-copy undistillable negative partial transposed states with the latter covering the vast majority of the surface of the separable polytope. 
\end{abstract}
\maketitle

\section{Introduction}

Best separable approximation (BSA) \cite{Sanpera} provides a convex decomposition of a given mixed quantum state which  fully unveils its  entanglement properties  \cite{Iha}. More recently, an  algorithm has been devised \cite{essent} for numerically efficiently achieving this decomposition and
in this paper we employ it  in order to study the  entanglement characteristics 
of random ensembles of two-qutrit states. The main focus of our studies is on   states   which
 are entangled  but with positive partial transposed (PPT) counterparts  \cite{Horodeckis98, Peres96, Horodeckis96} and in consequence undistillable/bound \cite{ BennettBrassard96,Horodeckis97,Rains1999a,Rains1999b,DuerBruss2000,Rains2001}. In this work we signify the PPT bound entangled states simply as bound entangled (BE) states.
Throughout the years, many important results have come to shed light on the intriguing class of BE states and by now  
many different  classes of BE states \cite{Horodeckis98,BennettMor99,DiVincenzoMor2003,HuberLami18,Sentis2016,Oh, Piani} have been identified, 
the robustness of the classes has been proven \cite{Ghosh}, tests for their detection \cite{Jens}  have been proposed, their non-local character \cite{Brunner},   steerability  properties \cite{Moroder} and  roles as activators \cite{activ} have been revealed. In addition to the studies on BE states, there are  efforts for identifying undistillable/bound entangled states with negative partial transposed (NPT) counterpart \cite{Horodecki0, DiVincenzo2000, Cirac, Watrous, Clarisse}. For this case though  undistillability  not stemming from the  PPT property, needs to be checked for an infinite number of copies of the state. The difficulty of  the latter task is vast and yet there is no safe conclusion on the existence of such states. In this paper, we also approach this subject by studying numerically one-copy  undistillable  (POCU) NPT  states.

The aim of this work is to add  more knowledge on the BE class of states. Using  BSA  and some straightforward theoretical tools we first propose  methods for creating random representative one-parametric families of BE. These methods permit us to generate   a number of random families for two-qutrit system and  draw conclusions on their characteristics. 
As expected, these families  are found to be deposed on the surface of separable polytope and are parts of bigger `island' formations.   We provide 
 some estimates on the average depth of these formations as well as on 
 their frequency of appearance on the surface which are in agreement with previous results  \cite{Jens}. We undertake a similar procedure for generating families of POCU  states and we conclude that these form a layer over the surface
of separable polytope that covers up the `islands' of BE states. From our studies we  also  conclude that the volume of BE states is  finite but negligible for the two-qutrit systems and that the PPT criterion for entanglement detection is accurate in very high degree.

The structure of the paper is the following. In Secs.~\ref{I} and \ref{Ib} our  means for  theoretical and numerical analysis are introduced:  we revise BSA, derive some useful related lemmas
for generating random families of BE states, and give details on the sampling methods for density matrices and on the algorithm for BSA analysis. In Sec.~\ref{II} we present numerical results on two-qubits states which form a ground of comparison for  higher dimensional systems.  The main numerical outcomes concern
two-qutrit states and these are found in Sec.~\ref{III}. In the last section, Sec.~\ref{IV},  we discuss questions  emerging from the theoretical methods and numerical results  of the current paper.

\section{ Families of BE states emerging from BSA  \label{I}}

BSA    was introduced in \cite{Sanpera} and its uniqueness was proven  using the convex property  of density matrices. In \cite{essent} the representation was re-introduced underlying its geometric aspects which in turn lead to  an efficient algorithm for its numerical realization. Here we  follow the notation and notions introduced in \cite{essent} -- which are in consistency with those in \cite{Sanpera}.

We start by providing some elements on the  geometric aspects of  mixed states which  help in introducing BSA. Let us adopt the usual Bloch-sphere representation and consider  pure states as unit vectors on its surface. For Hilbert space dimension  $N>2$, physical states occupy only a sub-manifold of the surface of the hyper-Bloch sphere \cite{essent} and any convex combination of these, lies inside this sub-manifold forming the \textit{convex hull body} of  mixed states.  Separable states are  states which can be written as convex combinations of pure product states and thus the separable convex set of states forms a \textit{separable polytope}  inside the convex hull body. States with sufficiently small  length  of Bloch vector  are necessarily inside the  \cite{volume1,Aubrun,Gurvits} and thus separable. The length of the Bloch vector of a density matrix is  related to its  purity $\mathrm{Tr} \rho^2$ but in this paper, we find it more convenient  use the inverse quantity:  the participation ratio $R=1/\mathrm{Tr} \rho^2$ \cite{volume1} with maximum value  $N$  and minimum value $1$ for pure states.
Also,   for our analysis  we are in need of  a measure of distance between two density matrices $\hat{\rho}$ and $\hat{\rho}'$ and for this we employ the measure of Hilbert-Schmidt distance: \begin{equation} D_{HS}\left(\hat{\rho}-\hat{\rho}'\right) =\sqrt{\mathrm{tr}\left[\left(\hat{\rho}-\hat{\rho}'\right)^2\right]}.\end{equation}

Given a density matrix $\widehat{\rho }$ describing a mixed multipartite quantum state, BSA is a unique convex decomposition over the separable polytope and  set of entangled states ($=$ convex hull body $/$separable polytope): 
\begin{equation}\widehat{\rho }=(1-B) \widehat{\rho }_{\mathrm{sep}}+B \widehat{\rho }_{\mathrm{ent}}.\label{one}\end{equation} 
In (\ref{one}) $\widehat{\rho }_{\mathrm{sep}}$  is, what we call in this work, the  \textsl{separable component}, $\widehat{\rho }_{\mathrm{ent}}$ the 
\textsl{essentially entangled} part  which cannot have any separable states as components, $B$ is a positive number in the range $\left[0,1\right]$ and $ \mathrm{Tr}\left[\widehat{\rho }_{\mathrm{sep}}\right]= \mathrm{Tr}\left[\widehat{\rho }_{\mathrm{ent}}\right]=1$.

The uniqueness of  BSA (\ref{one})  is justified from the fact that among all possible convex decompositions of  a state $\widehat{\rho }$ over the convex hull body of entangled states and convex polytope separable states,   the positive number $B$ attains its minimum value. Since the decomposition is  a result of extremization, both $\widehat{\rho }_{\mathrm{sep}}$ and $ \widehat{\rho }_{\mathrm{ent}}$ are  states on the boundaries of their representative sets. The essentially entangled component is on the  surface of the convex hull body  separating positive Hermitian matrices from non-positive ones and therefore of reduced rank.  The \textit{rank theorem} provides upper bounds on the rank of the essentially entangled component. This has been derived in \cite{Sanpera} for the bipartite case and extended in \cite{essent} for the general case:
The maximum rank $d_{E\max }$ of an essentially entangled component $\widehat{\rho }_{\mathrm{ent}}$ for a system of dimension $N$ composed by $K$ subsystems each of  them of dimension $N_k$, is $N-\sum_{k=1}^KN_k+K-1$. A direct consequence of this theorem is that for $N=4$ and $N=6$ the rank of $\widehat{\rho }_{\mathrm{ent}}$ is  one.

Respectively, the separable component lies on the surface of the polytope of separable density matrices, separating separable from entangled states. Since  both entangled and separable states are positive matrices, the rank of the separable component is not   reduced in the general case. On the other hand   under the PPT operation   $\widehat{\rho }_{\mathrm{sep}}$ is often mapped on the borders of positive with non-positive Hermitian matrices and therefore its rank is  reduced. As we explain further in this session this phenomenon always holds true  when $N=4$ or $6$.

\medskip

Let us list here some straightforward statements about the decomposition (\ref{one}). 
For this purpose we apply the operation of partial transposition  to left and right-hand sides of (\ref{one}),  obtaining the partial transposed version of it:
\begin{equation}\widehat{\rho }^{\Gamma}= (1-B)\widehat{\rho }_{\mathrm{sep}}^\Gamma+B \widehat{\rho }_{\mathrm{ent}}^\Gamma ,\label{too}\end{equation}
where $\widehat{\rho }_{\mathrm{sep}}^\Gamma$ necessarily a separable  state and $\widehat{\rho }_{\mathrm{ent}}^\Gamma $ a  Hermitian operator that is non-positive $\widehat{\rho }_{\mathrm{ent}}^\Gamma \ngeq 0$ when $\widehat{\rho }_{\mathrm{ent}} $ is of rank one (see Appendix \ref{A}). There we also conjecture that  this  property  holds  true in bipartite systems even if $\widehat{\rho }_{\mathrm{ent}}$ is of rank higher than one. 

\begin{enumerate}
	\item A state $\widehat{\rho }$ is \textsl{separable} \textsl{iff} $B=0$ in (\ref{one}).
	\item A state $\widehat{\rho }$ is \textsl{entangled} \textsl{iff} $B>0$ in (\ref{one}).
	\item A state $\widehat{\rho }$ is a PPT state \textsl{iff} $\widehat{\rho }^{\Gamma}\geq 0$ in (\ref{too}).
	\item A state $\widehat{\rho }$ is a NPT state  \textsl{iff} $\widehat{\rho }^{\Gamma}\ngeq0$ in (\ref{too}).
	\item A state $\widehat{\rho }$ is a BE state \textsl{iff} $\widehat{\rho }^{\Gamma}\geq 0$ in (\ref{too}) and $B>0$  in (\ref{one}).
	\item  Given the BSA (\ref{one}) for a density matrix $\widehat{\rho }$ and with $B\neq 0$ then
 \begin{equation}(1-B') \widehat{\rho }_{\mathrm{sep}}+B' \widehat{\rho }_{\mathrm{ent}}.\label{tri}\end{equation} 
with $1 \geq B'>0$ and $B'\neq B$ is the BSA for another density matrix $\widehat{\rho' }$, the one equal to (\ref{tri}).
\end{enumerate}

The last statement, a simple consequence of  convexity of the sets involved, permits us to start with an entangled state $\widehat{\rho }$  
with known BSA (\ref{one})  as generator/seed,  and create a one-parametric family of entangled states  
\begin{equation} f\left(\widehat{\rho },\beta\right)=(1-\beta) \widehat{\rho }_{\mathrm{sep}}+\beta \widehat{\rho }_{\mathrm{ent}}~~,\label{f}\end{equation}
with the parameter $\beta$ taking values in the range $\left[0,1\right]$. 

If in (\ref{f}) one employs as seed  a BE state   then a \textsl{family
of BE states} can be created for $\beta:\left[0,B_C\right]$ where $B_C$ is a critical value of the weight which can be  identified
by partially transposing (\ref{f}) and setting this to zero. This critical weight defines a state $\widehat{\rho }_C$ 
\begin{equation}\widehat{\rho }_C=(1-B_C)\widehat{\rho }_{\mathrm{sep}}+B_C \widehat{\rho }_{\mathrm{ent}}~~,\label{BC} \end{equation}
on the boundary of the BE family with NPT states. Then simply $\delta=D_{HS}\left(\widehat{\rho }_C -\widehat{\rho }_{\mathrm{sep}}\right)$
 quantifies   the `depth' of this BE family. We proceed with two  Lemmas which provide an alternative way for constructing families of BE states using as seeds NPT states  with specific properties.

\begin{itemize}

\item \textbf{Lemma 1.} An NPT entangled state (\ref{one})   with  $\widehat{\rho }_{\mathrm{sep}}^\Gamma$ in  (\ref{too}) being of full rank,  gives rise to a family of BE states  $f\left(\widehat{\rho },B\rightarrow 0\right) $.

\item \textbf{Lemma 2.} An NPT entangled state (\ref{one}) with  $\widehat{\rho }_{\mathrm{sep}}^\Gamma$ in  (\ref{too}) being of reduced rank, gives rise to a family of BE states  $f\left(\widehat{\rho },B\rightarrow 0\right) $ if  there is no  eigenvector  $\left|\phi\right\rangle$ in the null eigenvalues subspace of  $\widehat{\rho }_{\mathrm{sep}}^\Gamma$ such that $\left\langle \phi\right| \widehat{\rho }_{\mathrm{ent}}^\Gamma \left|\phi\right\rangle \ngeq 0$.

\end{itemize}
The proofs of the Lemmas  can be found in  Appendix \ref{B}.
A direct consequence of  Peres-Hordecki \cite{Peres96,Horodeckis96}
 criterion and Lemma 1 is that for $N=4$ and $6$ the $\widehat{\rho }_{\mathrm{sep}}^\Gamma$ in  (\ref{too}) is of reduced rank and therefore this lies on the borders between separable states and non-positive Hermitian operators. In addition, for such dimensions there is no NPT states satisfying  Lemma 2. We numerically confirm these  statements in  Section~\ref{II}.

\vspace{0.5cm}

 In Sec.~\ref{III}  we  employ the Lemmas $1$ and $2$  to identify  NPT states which can  serve  as seeds of one-parametric families of BE states $f\left(\widehat{\rho },\beta\right)$ for $\beta:\left[0,B_C\right]$.  The families of BE states created this way  can be considered as representative ones and analysis on them  provides estimates on the average characteristics  of  BE formations.

\section{On the algorithm performing BSA and  sampling methods \label{Ib}}

In this paper we   perform numerical analysis and it is important to give  information on the algorithm performing BSA, as well as, on the sampling methods for the density matrices.

 Concerning the algorithm, we closely follow the procedure prescribed in  \cite{essent} achieving an accuracy $\Delta B$ of the order $10^{-4}$  on the estimation of the extremum weight  $B$ in (\ref{one}) for two-qutrit states. This accuracy signifies that  if a state is analyzed by the algorithm more than once, the deviation of the results is of this order or less. In \cite{essent} is prescribed that the number of random vectors sampling the convex space at each step of the method should be of the order $N^4$ where $N$ the dimension of the Hilbert space of the combined system. We have found out that this number should be increased by a factor $\lambda$ for an accuracy consistent with the desired accuracy. The BSA algorithm in addition to $B$ provides the two density matrices components (\ref{one}) and the accuracy $\Delta \epsilon$ in their  spectrum is critical for the analysis performed in this paper. By analyzing many times several states and also with the use of the rank theorem, we draw conclusions on $\Delta \epsilon$ and most importantly we provide sufficient  thresholds $T_0$ for non-vanishing eigenvalues. Our conclusions  are summarized in the Table~\ref{TII} of  Appendix \ref{C} where we  also provide more details on the programs. Finally in our analysis we have excluded states  with one or more eigenvalue less than $10^{-4}$ since the algorithm is applicable only to full rank states. For these cases the algorithm suggested in \cite{Love} appears as good alternative to our method --especially when  $\widehat{\rho }_{\mathrm{ent}}$ is of rank one.

While for generating random  pure states   the generally admitted Haar measure exists, sampling mixed states is an intriguing subject with different options \cite{Karol1,Karol2}  and ongoing research.
In this paper we have employed two independent methods for sampling our space since the question on the dependence of the volume of BE states on the measure
is also a question of potential interest (see \cite{Slater1} and references therein for more intense studies on this subject). Our first choice is   the \textit{flat measure} --as we call it in this paper, which has been used
in the very first works on the volume of separable \cite{volume1} and BE states \cite{ volume2}. More specifically,  the eigenvalues and eigenvectors of a density matrix are treated  as two different sets; the first are sampled uniformly over its geometric space (simplex) and the second  use unitary projectors uniformly sampled over the Haar Measure. The first step is straightforward using the instructions in the Appendix of  \cite{volume1}, for the second step we have used random unitary matrices produced according to the procedure \cite{Karol3} and relevant programs \cite{Poland}.
The second sampling method that we use is the \textit{induced measure}  \cite{Karol1} which has gained a lot of attention the last years. This measure  follows the more natural procedure  using ancillary systems and tracing out degrees of freedom of pure random states of higher dimensions. The dimension of the ancillary system defines the measure and metric of the produced random ensemble. If the dimension is the same as for the system one obtains a Hilbert-Schmidt ensemble, while if the dimension is higher one obtains induced measure ensembles  (with Bures ensemble as a subcase). We have found out that each of these ensembles is circumscribed normally distributed around a certain degree of participation ratio $R$, so in order to obtain a distribution of states which covers the full range of  $R$ we use different dimensions for the ancillary system
and also a method suggested in \cite{Karol1} that includes projection onto maximally entangled states for the ancillary system (instead of tracing out). For our programs we have used Ginibre matrices generated by the Mathematica package \cite{Poland}. 
In  Appendix \ref{DM} we present graphically distributions of two-qubit and three-qubit density matrices according to flat and (combined) induced measure.

\section{ BSA on two-qubit  states \label{II}}

 According to the Peres-Horodecki criterion \cite{Peres96,Horodeckis96},  two-qubit systems cannot accommodate BE states. We think though that it is worth  analyzing this case since the results provide  information on the  BSA properties for this system which has not been reported elsewhere. We perform BSA  on $1200$ randomly sampled NPT density matrices, $600$ according to flat measure and $600$  according to induced measure and in  Fig.~\ref{fig3} we summarize the joint outcomes.

The rank theorem \cite{Sanpera} applied on a two-qubit  system, dictates that $\widehat{\rho }_{\mathrm{ent}}$ is necessarily  of rank one and therefore  (\ref{one}) attains the simpler form:
\begin{equation}\widehat{\rho }=(1-B) \widehat{\rho }_{\mathrm{sep}}+B \left| \psi_{ent}\right\rangle\left\langle \psi_{ent}\right|.\label{for}\end{equation} 
Our numerical results confirm  Eq. (\ref{for}) and as it would be expected,    the weight $B$ in Fig~\ref{fig3}~(a) in average drops with the decrease of purity.
For $R>3$ confirming previous results \cite{volume1,Gurvits},   no entangled states are detected. 
Interestingly   in Fig~\ref{fig3}(b) we observe that the concurrence \cite{concu} of  $\left| \psi_{ent}\right\rangle$ in (\ref{for}), on average is  increasing with $R$, and that  this is a maximally entangled state for a big fraction of  density matrices ($\approx 72\%$ for our data).
In the Appendix \ref{F} we show how the latter observation can be employed  for enhancing the distillation procedure of two-qubit NPT states.

Confirming Peres-Horodecki criterion we see in Fig.~\ref{fig3}~(c)-(d), that none of the conditions of Lemmas $1 -2$, which could lead to families of  bound states from NPT states are fulfilled. More
specifically,    $\rho_{sep}^{\Gamma}$ is of reduced rank (below the threshold $T_0$, see Appendix \ref{C}) and the
overlap of the null eigenvector $\left| \phi_0\right\rangle$ of $\rho_{sep}^{\Gamma}$ with the $\rho_{ent}^{\Gamma}=\left| \psi_{ent}\right\rangle\left\langle \psi_{ent}\right|$  always negative. Finally, we have not observed any differentiation on the results  between flat and induced measure sampling.
\begin{widetext}
\begin{center}
\begin{figure*}[h]{\centering{\includegraphics*[width=0.85\textwidth]{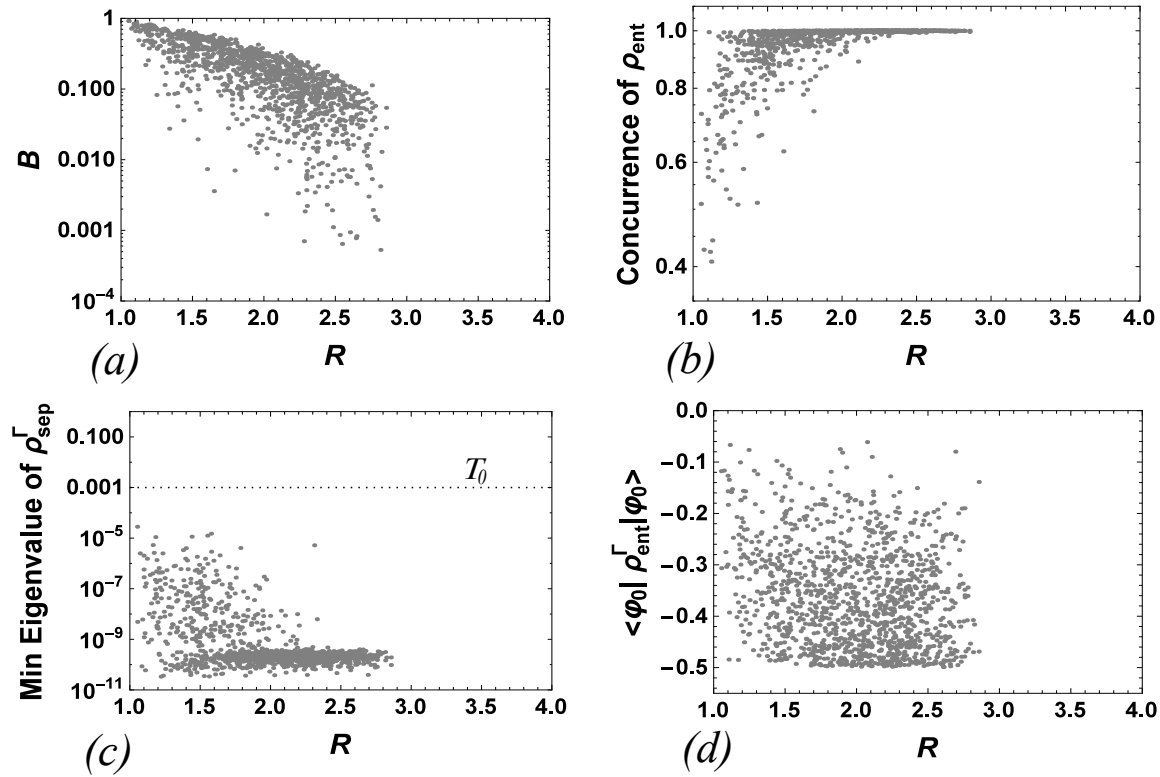}}}\caption{BSA  analysis on $1200$ NPT two-qubit random states ($600$ flat $+$ $600$ induced measure). \textit{(a)} The weight $B$ over participation ratio, \textit{(b)} the concurrence of $\widehat{\rho }_{\mathrm{ent}}=\left| \psi_{ent}\right\rangle\left\langle \psi_{ent}\right|$, \textit{(c)} the minimum eigenvalue of the partial transposed separable component $\widehat{\rho}_{sep}^{\Gamma}$, \textit{(d)} the overlap of the $0$ eigenvalue eigenvector $\left| \phi_0\right\rangle$ of $\widehat{\rho}_{sep}^{\Gamma}$, with the $\widehat{\rho}_{ent}^{\Gamma}$.
The error on the estimation of the weight $B$ is less than $\Delta B= 5 \times 10^{-4}$ while the threshold for non-vanishing eigenvalue $T_0=10^{-3}$, (see Appendix \ref{C}).} \label{fig3}\end{figure*}
\end{center}
\end{widetext}

\section{ BSA on two-qutrit states \label{III}}

\begin{figure}[h] \includegraphics[width=0.45\textwidth]{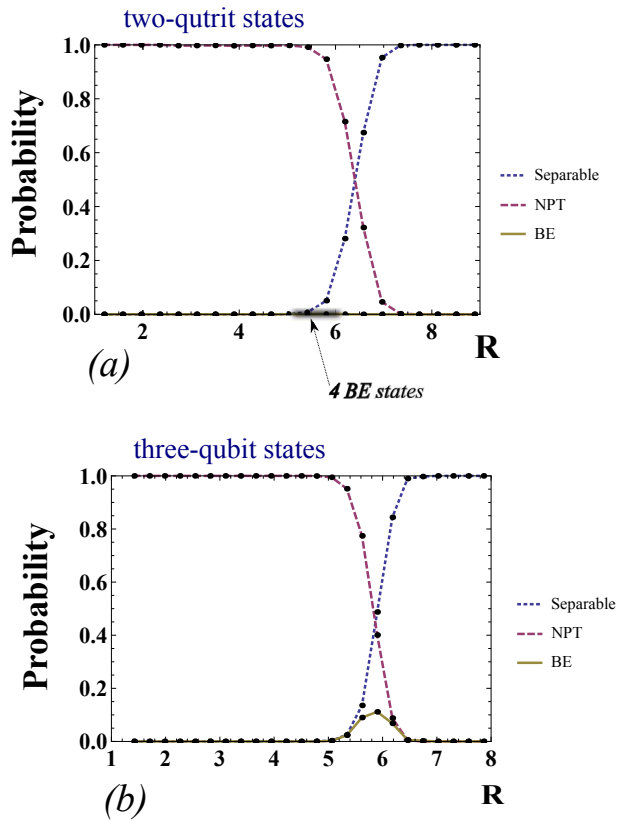}
\vspace{1cm}
\caption{The probability for  separable/NPT/BE states to occur in a random   ensemble   of \textsl{(a)} two-qutrit  and  \textit{(b)} three-qubit systems, as a function of the participation ratio.  The graphs are based on statistics on $30000$ and $5000$ states respectively, sampled according to flat measure. In graph \textsl{(a)} the detected BE states  are too few (see Table~I) for their probability amplitude to be visible. } \label{fig2}\end{figure}

We start by performing BSA on PPT states in order to obtain an estimate on the BE states' volume as compared to the NPT's and separable ones'.
We have tested $30000$ random  states (among which $3392$ PPT) sampled according to flat measure and identified only $4$  BE states.
The results are summarized in Fig.~\ref{fig2}~(a), where it becomes obvious that the volume of BE states is  negligible (but finite in accordance with \cite{Aubrun}),  below the $0.1$\% of total volume.
In Fig.~\ref{fig2}~(b) we  present the corresponding graph for the three-qubit system testing $5000$ random states sampled over flat measure. Fig.~\ref{fig2}~(b)  reproduces   Fig.~$7$ in \cite{ volume2} with more accurate methods and gives a volume of BE states approximately  $2$\% of the total volume. This comparative study  on composite systems of similar dimensions confirms the known fact \cite{Aubrun} that the tensor product structure of the Hilbert space does matter for bound entanglement.

 In  Table $I$ we list the properties of the  few identified BE states: $R$, $B$ weight, the rank of  $\widehat{\rho }_{\mathrm{ent}}$,  the minimum eigenvalue of the partial transposed separable component $\widehat{\rho}_{sep}^{\Gamma}$ (see Lemma 1),  and the overlap of the minimum eigenvalue eigenvector $\left| \phi_0\right\rangle$ of $\widehat{\rho}_{sep}^{\Gamma}$ with  $\widehat{\rho}_{ent}^{\Gamma}$ (see Lemma 2). We also employ BE states as seeds for BE families and calculate  the critical weight $B_C$ (\ref{BC}) of the latter.

\begin{table}[thb]
    \centering
		\caption{ BE states detected.}
    \begin{tabular}{|c|c|c|c|c|c|}
        \hline
    &   Weight  & Rank of                         & Min Eigenvalue             &      Overlap     &     \\ 
	$R$		&	 $B$   & $\widehat{\rho}_{ent}^{\Gamma}$ & of $\widehat{\rho}_{sep}^{\Gamma}$ &  $\left\langle \phi_0\right| \widehat{\rho }_{\mathrm{ent}}^\Gamma \left|\phi_0\right\rangle   $      &$B_C$     \\ \hline 
	$5.16$  &$0.0151$&$1$&$0.0033$ &$-0.091$&$0.0161$\\	
	$5.75$ &$0.0014$ &$1$&$0.0008$ &$-0.235$&$0.0031$\\	
	$5.88$  &$0.0036$&$1$ &$0.0033$&$-0.325$&$0.0089$\\ 
	$6.11$ &$0.0046$&$1$&$0.0068$ &$-0.275$ & $0.0214$\\\hline			  
              
    \end{tabular}
\end{table}

\pagebreak

We proceed by  performing   BSA on NPT states. This  a 
time-consuming numerical task and  we have  analyzed  via BSA $600$ random NPT states --$300$ for flat and $300$ for the induced measure.  
 We present  the results in Fig.~\ref{fig5} and we observe that as for the two-qubit systems (see Fig.~\ref{fig3}), the weight $B$ in average drops  with participation ratio.   In  Fig.~\ref{fig5}~(b) the distribution on rank of the essentially entangled component, $\widehat{\rho}_{ent}$ of the random  states is presented and as predicted by the rank theorem this does not exceed the value $5$. 

In Fig.~\ref{fig5}~(c), we now see that  NPT states appear   above the secure  threshold $T_0$ for non-vanishing eigenvalues (see Appendix \ref{C}) thus fulfilling the criteria of Lemma $1$ with certainty. These states can serve as  seed states able to  generate    BE families of states. Similarly we detect some states fulfilling the criteria of Lemma $2$. Finally, few of the states fulfill the criteria of both Lemmas.
In overall, we have detected $40$ NPT states  which can generate BE families  and since the threshold $T_0$ is  sufficient
but not necessary for their detection,  we may conclude  that the surface of separable polytope is covered by BE states at a  considerable rate  
$>10$\%. On the other hand, from our results we have not been able to draw safe conclusions
on  the differentiation of results due to different samplings of the states under analysis.

In the next step, we use the detected  seed NPT states to generate BE families and estimate the corresponding 
 critical weight $B_C$ (\ref{BC}) and  depth $\delta$.
The results are summarized in   Fig.~\ref{fig6}~(a) and (c). Combining the rate of occurrence of NPT seed states with the average depth of the families we  confirm the results of analysis on PPT states; the volume of BE states does not exceed $1/1000$ of the total volume of states.

The identified BE families $f$ are not isolated and these are part of bigger formations, `islands' of BE states which are deposed on the surface
of the separable polytope, see Fig.~\ref{fig6}~(d). This phenomenon is the subject of \cite{Ghosh} and more recently jagged islands of BE states have been constructed in \cite{Slater2}.  From the point of view of BSA (\ref{one}) it is evident that infinitesimal deviations on a seed NPT state $\widehat{\rho }$ will give infinitesimal deviations to its components  and in turn  BE families  in the neighborhood of the initial one.

\subsection{Generating families of  POCU states}

We proceed with a closely related matter to PPT bound entanglement, the one of undistillability of NPT states, and we examine whether  random
families of POCU states can be constructed  in a similar fashion as for BE states. The answer is positive and the numerical
analysis gives evidence that  in their vast majority ($100\%$ of tested cases) NPT states  generate  one-parametric families of  POCU states as $f\left(\widehat{\rho },B\rightarrow 0\right)$. To estimate the depth $\sigma$ of  these POCU families  we follow similar steps as for the BE families: we start with the BSA of a random NPT state $\widehat{\rho }$ (\ref{one}) and we decrease $B$ in (\ref{f}) until we identify numerically the critical weight $B_c$ such that  $\widehat{\rho }_c$
\begin{equation}\widehat{\rho }_c=(1-B_c)\widehat{\rho }_{\mathrm{sep}}+B_c \widehat{\rho }_{\mathrm{ent}}, \end{equation}
 is a POCU state. The states   $f\left(\widehat{\rho },B\right)$, with $0<B<B_c$ comprise a family of POCU states. Then $\sigma=D_{HS}\left(\widehat{\rho }_c -\widehat{\rho }_{\mathrm{sep}}\right)$
provides the depth of the POCU family. In the Fig.~\ref{fig6}~(b) we present the distribution of $\sigma$ for $200$ POCU families which have been created using as generators random $NPT$ states sampled according to flat measure.

BE states are undistillable for any number of copies and  further analysis on the BE families constructed in the previous section shows the expected phenomenon:  BE families are subsets of   POCU families. To numerically prove this, we use the detected NPT seed states for BE families,  as generators for POCU families and  we compare the critical weights $B_C$ and $B_c$. The (sorted) results  are presented graphically in Fig.~\ref{fig6}~(c) where one may observe that $B_C<B_c$ holds in all cases under test. In a similar way,  $\delta<\sigma$ for all tested NPT seed states.

Finally a few words about the algorithm that we employ to check numerically the one-copy undistillability of states. This relies on the Lemma 2 of \cite{DiVincenzo2000}
and according to it, it is sufficient to check whether projections $\widehat{\rho}_{2\otimes 3}=P_A\otimes 1_B \widehat{\rho} P_A^{\dagger}\otimes 1_B$ on the density matrix result in
$\widehat{\rho}_{2\otimes 3}^{\Gamma} \ngeq 0$. If the latter holds true we conclude that the state is distillable while if all random ($50.000$) projections
that we try out give $\widehat{\rho}_{2\otimes 3}^{\Gamma} > 0$ we characterize the state as POCU state. To parametrize the component $\left|a_0\right\rangle$  of the  projector $P_A=\left|a_0\right\rangle\left\langle a_0\right|+\left|a_1\right\rangle\left\langle a_1\right|$ with $\left\langle a_0 \right|\left. a_1\right\rangle=0$ we use the convenient representation from \cite{Klimov} for a qutrit state in terms of angles,
\begin{eqnarray}
\left|a_0\right\rangle= & \mathrm{sin}\left(\xi/2\right) \mathrm{cos} \left(\theta/2\right) \left|1\right\rangle+ e^{i \phi_{12}} \mathrm{sin}\left(\xi/2\right) \mathrm{sin} \left(\theta/2\right) \left|2\right\rangle  \nonumber\\
& + e^{i \phi_{13}} \mathrm{cos}\left(\xi/2\right)  \left|3\right\rangle  \label{Kli}
\end{eqnarray}
 which permits us to have also an analytic expression for $\left|a_1\right\rangle$. The random search on the projectors is performed as random search on the angles of $\left|a_0\right\rangle$ (\ref{Kli}) and on the free angle parameters of $\left|a_1\right\rangle$.

\begin{widetext}
\begin{center}

\begin{figure*}[h]{\centering{\includegraphics*[width=0.85\textwidth]{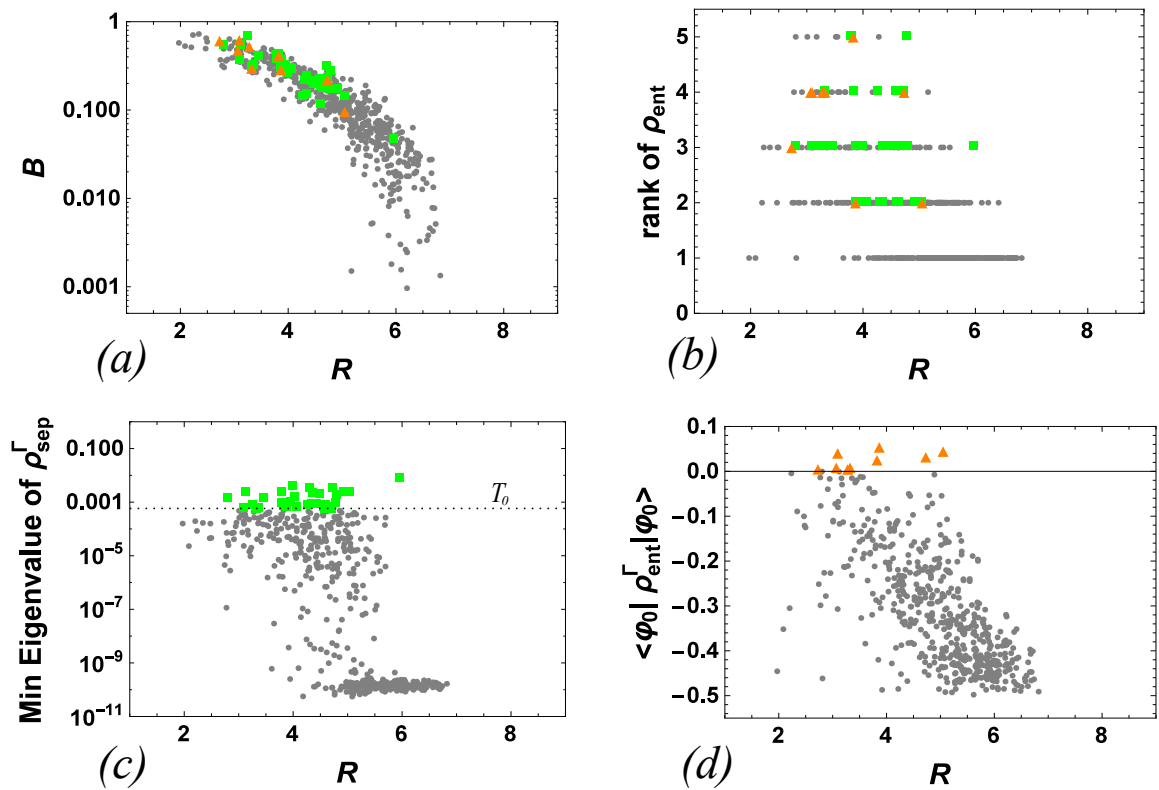}}}\caption{BSA  on $600$ two-qutrit NPT random states ($300$ flat $+$ $300$ induced measure).  The states which can act as generators  for BE families according to Lemma 1 are marked as green squares, and those according  to Lemma 2 as orange triangles.    \textit{(a)} The weight $B$ over participation ratio, \textit{(b)} the rank of  $\widehat{\rho }_{\mathrm{ent}}$, \textit{(c)} the minimum eigenvalue of the partial transposed separable component $\widehat{\rho}_{sep}^{\Gamma}$, \textit{(d)} the overlap of the minimum eigenvalue eigenvector $\left| \phi_0\right\rangle$ of $\widehat{\rho}_{sep}^{\Gamma}$, with the $\widehat{\rho}_{ent}^{\Gamma}$. The error on the estimation of the weight $B$ is less than $\Delta B= 5 \times 10^{-5}$ while the sufficient threshold for non-vanishing eigenvalue $T_0=5 \times 10^{-4}$, see Appendix \ref{C}.} \label{fig5}\end{figure*}

\begin{figure}[h]\includegraphics*[width=0.85\textwidth]{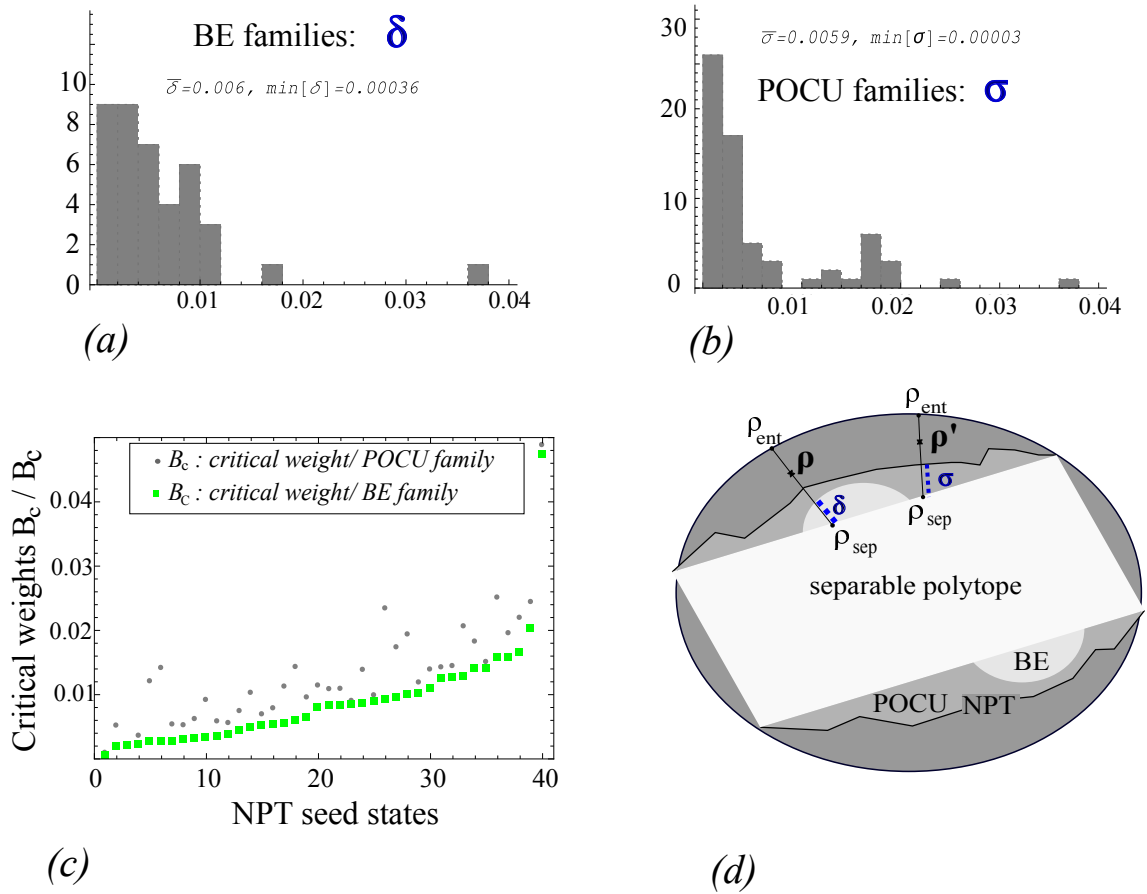}\caption{\textsl{(a)} Distribution of depth $\delta$ for  BE families  generated via  $40$  NPT seed states  (see Fig.~\ref{fig5}). \textsl{(b)} Distribution of  $\sigma$ for  POCU families of states generated by $200$  NPT states (randomly sampled according to flat measure). \textsl{(c)} The critical weight $B_C$ for the BE families of states generated via  $40$ NPT seed states and the corresponding critical weight $B_c$ for the  POCU families generated by the same  states. \textsl{(d)} A very schematic representation of the separable polytope, BE islands,  the layer of POCU states,  NPT seed states $\rho$ acting as generators for  BE families and  NPT states $\rho'$ generating  POCU families.} \label{fig6}\end{figure}
\end{center}

\end{widetext}

\section{Discussion\label{IV}}

Summarizing the main analytical and numerical findings of this paper, one first useful conclusion is that random one-parametric BE families can be generated
at an efficient rate by performing BSA on NPT states. Applying the aforementioned technique we have been able to create a number of such families and draw some preliminary conclusions on their properties   and their placement in the convex space of density matrices; the families lie on the surface of the separable polytope  without much penetrating  the NPT volume. Then we have proposed a similar procedure for generating one-parametric families of POCU NPT states and we have observed that the latter form a layer over the surface of the separable polytope and covering  up the BE formations. Finally by applying BSA on a considerable number of PPT states we have seen that BE states  for two-qutrit systems are so rare that PPT criterion can be applied with confidence of more than  $99,9\%$. For our numerical studies
we have employed two independent sampling methods and our generic conclusions seem to hold for both methods. However in order to draw decisive conclusions on the sampling matter as well as for providing  precise  quantitative estimations on quantities under examination in this paper (such as $\delta$, $\sigma$),  a much higher number of BSA tests  is necessary.
	
A straightforward extension of this paper would be to perform BSA analysis on higher-dimensional bipartite systems and track the rate of growth
of the BE states' volume. The linear programming algorithm \cite{essent}  used in this work scales as $N^{12}$ where $N$ the dimensional of the total Hilbert space, and dimensions up to $N=16$ are still tractable if high computational power is available. Another possibility for reaching higher dimensions  is to use the current programs  to conclude on the BE detection ability of  more computationally accessible methods,  such as covariance matrix criteria \cite{Eisert}, and proceed with the latter. 

On the theoretical level a  finding of this paper deserving further investigation, is that
both (PPT) BE and (NPT) POCU states  are correlated with a BSA weight $B$  of the same (low)
 order of magnitude.  This phenomenon gives the suggestion   that undistillability of a state might be related  not only  with its character  under partial transposition but also with its BSA geometric properties. 
Along the same line,   the example in the Appendix \ref{F} gives  evidence that BSA can provide useful knowledge to a distillation process and we believe  that more results can be discovered on this subject.   Finally, in this work we use the Peres-Horodecki criterion to explain the fact that for $N=4$ and $6$ partial transposition always maps the separable component of a state to the borders of separable polytope and non-positive operators. An independent proof  which could be helpful for a better understanding of the geometric space of entangled states and related maps, is still pending.

\section*{Acknowledgement}
The authors are grateful to Andreas Osterloh, Jens Siewert, Karol ̇{\.Z}yczkowski  for  essential feedback and exchange. The authors acknowledge financial support from the Nazarbayev University ORAU grant ``Dissecting the collective dynamics of arrays of superconducting circuits and quantum metamaterials'' (no. SST2017031) and MES RK state-targeted program BR$05236454$. A.M. is also thankful to ICTP for warm hospitality and financial support.

\appendix
\section{On the non-positivity of the partially transposed essentially entangled component \label{A}}

	Let us consider bipartite symmetric systems of total dimension $N=N_K \times N_K$ with the assumption that $\widehat{\rho }_{\mathrm{ent}}$ in (\ref{one}) is of rank 1, or else  $\widehat{\rho }_{\mathrm{ent}}=\left|\psi\right\rangle\left\langle \psi \right|$.
	Let  the Schmidt decomposition for $\left|\psi\right\rangle$ be  $$\left|\psi\right\rangle=\sum_{i}^{N_K} \sqrt{\alpha_i} \left|b_i\right\rangle \left|c_i\right\rangle .$$  The essentially entangled component   in the aforementioned  basis is
	\begin{equation}
	\widehat{\rho }_{\mathrm{ent}}=\sum_{i j}^{N_K} \sqrt{\alpha_i \alpha_j} \left|b_i\right\rangle \left\langle b_j \right|\otimes\left|c_i\right\rangle \left\langle c_j \right|.
	\label{ho}\end{equation}
	and after partial transposition on the second system, \ref{ho} becomes
	\begin{equation}
	\widehat{\rho }_{\mathrm{ent}}^{\Gamma}=\sum_{i j}^{N_K} \sqrt{\alpha_i \alpha_j} \left|b_i\right\rangle \left\langle b_j \right|\otimes\left|c_j\right\rangle \left\langle c_i \right|.
	\end{equation}
	One can check that any vector $\frac{1}{\sqrt{2} }\left(\left|b_k\right\rangle \left|c_l\right\rangle -\left|b_l\right\rangle \left|c_k\right\rangle\right) $ with $k\neq l$ is
	an eigenvector with negative eigenvalue $-\sqrt{\alpha_k \alpha_l} $. Therefore the minimum dimension of the negative subspace of   $\widehat{\rho }_{\mathrm{ent}}^{\Gamma}$ is  $\frac{{N_K}!}{\left({N_K}-2\right)!2!}$.

\begin{figure}[h] \includegraphics[width=0.4\textwidth]{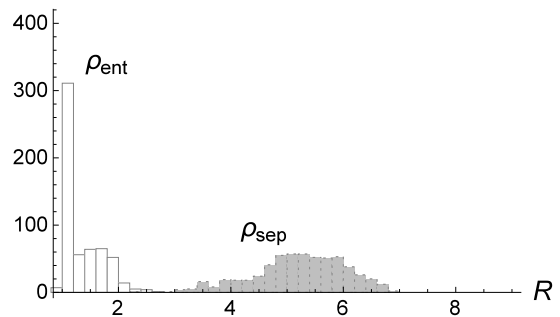}\caption{The distribution of $\widehat{\rho }_{\mathrm{ent}}$  and $\widehat{\rho }_{\mathrm{sep}}$ over participation ratio. For this graph
$600$ random NPT  states have been used, the same as for  Fig.~\ref{fig5}.} \label{fig7}\end{figure}

Now let us consider a higher rank $M<d_{E max}$ (see rank theorem in Sec.~\ref{I}) for $\widehat{\rho }_{\mathrm{ent}}=\sum_{m=1}^{M}\lambda_m \left|\psi_m\right\rangle \left\langle \psi_m \right|$. Every  eigenvector $\left|\psi_m\right\rangle$ of  $\widehat{\rho }_{\mathrm{ent}}$ after partial transposition transforms into a negative operator $\hat{V}_m$ negative subspace of dimension $\frac{{N_K}!}{\left({N_K}-2\right)!2!}$. However this does not necessarily imply that $\widehat{\rho }_{\mathrm{ent}}^{\Gamma}= \sum_{m=1}^{M}\lambda_m \hat{V}_m$ is negative; in the general case the probability that the
	different negative subspaces have some overlap (and in consequence $\widehat{\rho }_{\mathrm{ent}}^{\Gamma} \ngeq 0$) is obviously very small especially as $N$ is increasing.
	On the other hand what we have observed in the two-qutrit system (but also in previous studies \cite{essent})  that among the $M$ eigenvectors there is a principal one $\left|\psi_p\right\rangle$ with $\lambda_p>0.5$. This can be understood from the fact that the essentially entangled component is located towards the outer surface of the convex body and of relatively high purity. In Fig.~\ref{fig7} we exhibit the distribution of $\widehat{\rho }_{\mathrm{ent}}$ and $\widehat{\rho }_{\mathrm{sep}}$ over the participation ratio for two-qutrit random states that confirms this statement. We conjecture that this phenomenon persists in higher dimensions of bipartite systems, i.e., there is principal eigenvector whose non-positivity under partial transposition dominates, and in consequence $\widehat{\rho }_{\mathrm{ent}}^{\Gamma} \ngeq 0$.
	
\section{Proofs of Lemmas \label{B}}

Let us consider a NPT state $\widehat{\rho}$ with BSA (\ref{one}) and also their partially transposed counterparts $\widehat{\rho}^{\Gamma}$, (\ref{too}).
According to statement $6$ in Sec.~\ref{I} the state $\widehat{\rho}$ generates a family of states $f(\widehat{\rho}, B)$ as in (\ref{f}) and let us denote by $f(\widehat{\rho}, B)^{\Gamma}$ the partially transposed counterpart of the family.

By the definition of  NPT states, at least a vector $\left|\phi\right\rangle$ exists such that
$\left\langle \phi \right| \widehat{\rho}^{\Gamma}\left|\phi\right\rangle <0$, or employing (\ref{too}) such that
\begin{equation} (1-B)\left\langle \phi \right|\widehat{\rho }_{\mathrm{sep}}^\Gamma\left|\phi\right\rangle+B \left\langle \phi \right|\widehat{\rho }_{\mathrm{ent}}^\Gamma \left|\phi\right\rangle <0.\label{mtoo}\end{equation}
Since $\widehat{\rho }_{\mathrm{sep}}^\Gamma \geq 0$ and $\widehat{\rho }_{\mathrm{ent}}^\Gamma \ngeq  0$, it is convenient to introduce here parameters $\epsilon\geq0$, $\gamma >0$ and re-express  (\ref{mtoo}) as
\begin{equation} (1-B)\epsilon-B \gamma<0\label{metoo}\end{equation}
where
\begin{eqnarray}
\left\langle \phi \right|\widehat{\rho }_{\mathrm{sep}}^\Gamma\left|\phi\right\rangle & =\epsilon \\
\left\langle \phi \right|\widehat{\rho }_{\mathrm{ent}}^\Gamma \left|\phi\right\rangle & =-\gamma.
\end{eqnarray}

Taking cases on the rank of $\widehat{\rho }_{\mathrm{sep}}^\Gamma$: 
\begin{itemize}
	\item If $\widehat{\rho }_{\mathrm{sep}}^\Gamma$ has full rank, then this implies that is a strictly positive operator and  then for any vector $\left|\phi\right\rangle$, $\epsilon>0$. In consequence
	there is a positive parameter $B_\ast=\frac{\epsilon}{\epsilon+\gamma}$ (derived by (\ref{metoo})) such  that if $B<B_\ast$, $\left\langle \phi \right| \widehat{\rho}^{\Gamma}\left|\phi\right\rangle >0$. It is easy to check that under these conditions $ \left\langle \phi \right|f(\widehat{\rho}, B<B_\ast)^{\Gamma} \left|\phi\right\rangle >0 $
	and therefore searching among all $\left|\phi\right\rangle$ one can identify a global minimum (positive) value for $B_\ast$ that we denote as $B_C$. This implies that a family of BE states exists for $B<B_C$.
	
	\item If $\widehat{\rho }_{\mathrm{sep}}^\Gamma$ has at least one $0$ eigenvalue but the corresponding to this eigenvalue, eigenvector $\left|\phi_0\right\rangle$ gives $\left\langle \phi_0 \right|\widehat{\rho }_{\mathrm{ent}}^\Gamma \left|\phi_0\right\rangle>0$. For this vector 
	obviously $ \left\langle \phi_0 \right|f(\widehat{\rho}, B\rightarrow 0)^{\Gamma} \left|\phi_0\right\rangle >0 $. For any other vector $\epsilon>0$ holds and  there is a critical $B_\ast$ that $ \left\langle \phi \right|f(\widehat{\rho}, B<B_\ast)^{\Gamma} \left|\phi\right\rangle >0 $. Similarly to previous case, one can identify a global minimum (positive) value for $B_\ast$ and identify a family of BE states.
\end{itemize}

\section{On the accuracy limits of the programs \label{C}}

The algorithmic procedure  that we use  \cite{essent} for our programs consists of a random sampling of
 the convex hull body and convex polytope of separable states,
a linear programming routine (simplex method) that selects the vectors among the sampled ones which satisfy the constrains of BSA
and a `loop' to gradually converge to the unique solution. The simplex method is applied
 exactly and thus the final accuracy  depends on the number $M=\lambda N^4$ of vectors/states
which are used to sample the space at each step and on the prescribed accuracy of convergence $\epsilon_C$ in the loop.  In turn the running time has a dependence $M^3$ on the number of vectors and has approximately inverse dependence on $\epsilon_C$. In the programs constructed for this work we try to balance between accuracy and running time since we aim to perform some low-level statistics.
Taking into account all the aforementioned factors   we summarize in the Table II the parameters of the  programs used in this work together with their accuracy limits. The latters  have been estimated by testing many different states, multiple times each one of these, and also with the use of the rank theorem. The program used for the numerical analysis of two-qutrit states can be found on the page www.qubit.kz

\begin{table}[thb]
    \centering
		\caption{ Parameters in programs and accuracies achieved}
    \begin{tabular}{|l|c|c|}
        \hline
        \textbf{System}  & \textbf{ two-qubit} & \textbf{two-qutrit}  \\ 
				                  & $\left(N=4\right)$              & $\left(N=9\right)$ \\ \hline \hline
			   Number     &              &     \\
				of input vectors:    &             &     \\

         $M=\lambda N^4$     &     $\lambda=15$          &    $\lambda=4$  \\ \hline 
				 Accuracy     &             &    \\
				 of convergence      &              &    \\
			 in the loop:    &            &     \\
        $\epsilon_C$ &  $10^{-15}$   & $10^{-10}$   \\ \hline \hline
          Accuracy    &               &  \\ 
					on the weight:    &               &  \\ 
              $\Delta B$    &    $ 5 \times 10^{-4}$    &   $ 5 \times 10^{-5}$ \\ \hline
							 Accuracy    &               &  \\ 
					on  eigevalues    &               &  \\ 
					of  components of BSA:    &               &  \\ 
              $\Delta \epsilon$    &  $ 3 \times 10^{-3}$      &  $5 \times 10^{-4}$ \\ \hline
						Secure threshold    &               &  \\ 
					 for non-vanishing  &               &  \\ 
					eigenvalues $T_0$:       &  $ > 10^{-3}$        &  $ >  5 \times 10^{-4}$    \\ \hline
              
    \end{tabular} \label{TII}
\end{table}

\section{ Distributions of Density Matrices \label{DM}}
\begin{figure}[h] \includegraphics[width=0.38\textwidth]{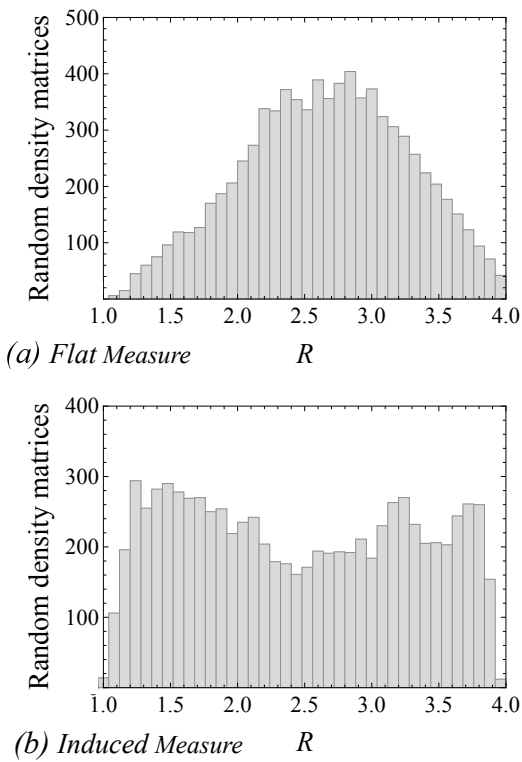}
\vspace{1cm}
\caption{Random ensembles of ($8000$) two-qubit density matrices  distributed over participation ratio $R$. Sampling according to \textit{(a)}  flat measure and \textit{(b)} induced measure using ancillary system of dimension $K=4, 8, 16, 40, 80 $ but also projection onto maximally entangled states (for four-partite and six-partite systems \cite{Karol1} in order to approach states of high purity).  } \label{fig1}\end{figure}

\begin{figure}[h] \includegraphics[width=0.38\textwidth]{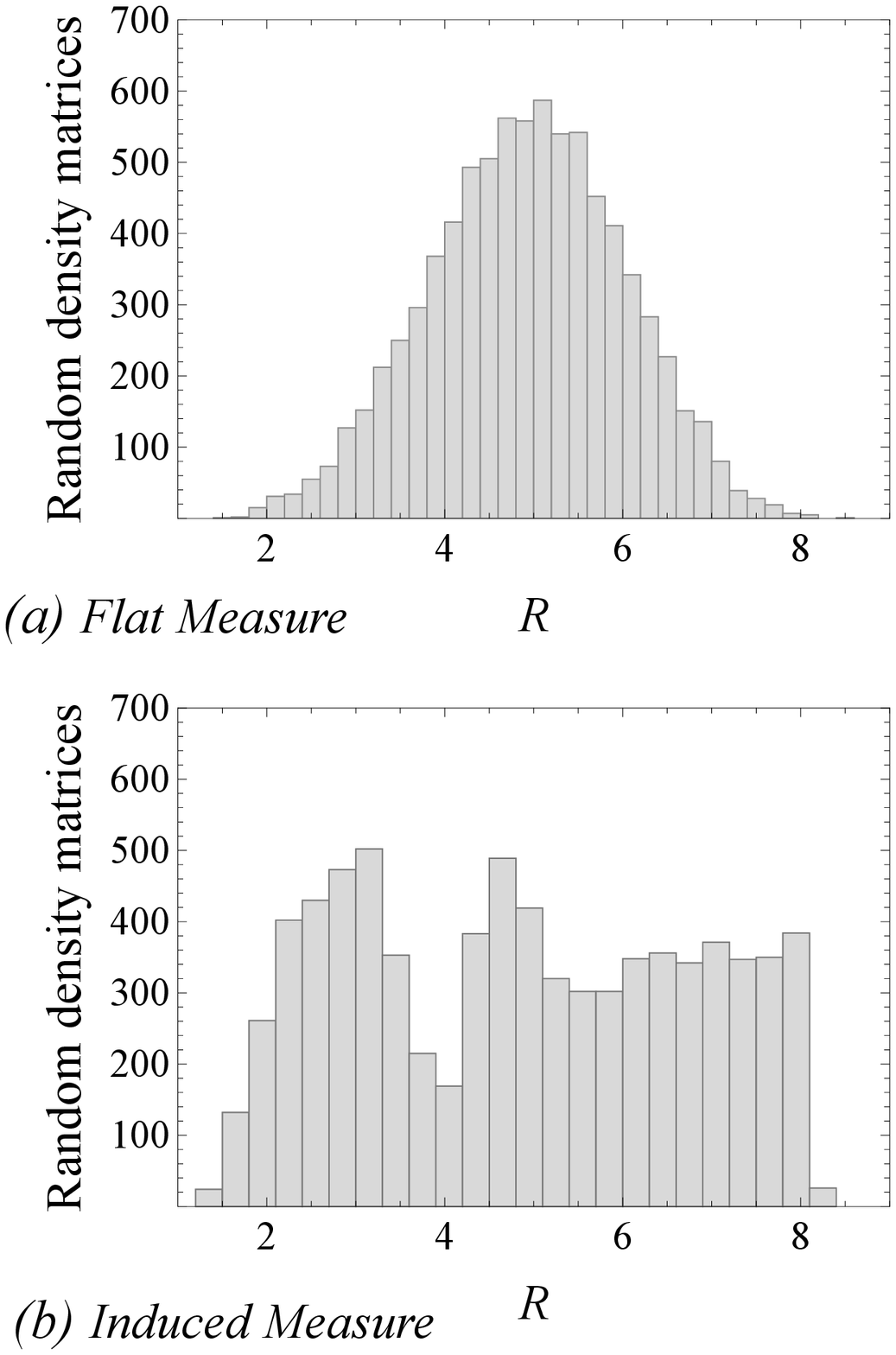}
\vspace{1cm}
\caption{Random ensembles of ($8000$) two-qutrit density matrices distributed over participation ratio. Sampling according to \textit{(a)}  flat measure and \textit{(b)} induced measure using ancillary system of dimension $K=9, 13, 18, 25, 36, 60 $ but also projection onto maximally entangled states  \cite{Karol1}.  } \label{fig1b}\end{figure}

In Figs.~\ref{fig1}-\ref{fig1b}   the distributions of  two-qubit and three-qubit density matrices over the participation ratio $R$  are presented for  sampling    according to  flat measure and to (combined) induced measure respectively.

\section{ Using BSA to skip the local filtering protocol for two-qubit states \label{F}}
All entangled two-qubit states are distillable \cite{Horodeckis97}, meaning that if a sufficient number of copies of the state are provided and local operations and classical communication are allowed, then at least an  EPR pair, $\left|\psi_{EPR}\right\rangle=1/\sqrt{2}\left(\left| 00\right\rangle+\left| 11\right\rangle\right)$ can be produced. If  the fidelity $F$ of the state $\hat{\rho}$ with the EPR pair, $F_{EPR}\left( \hat{\rho} \right)=\mathrm{Tr}\left[ \hat{\rho} \left|\psi_{EPR}\right\rangle \left\langle \psi_{EPR}\right| \right]$, is greater than $1/2$, distillation can be achieved via the recurrence protocol \cite{hash}, followed up by  the hashing protocol \cite{hash}. In the habitual case, where $F_{EPR}<1/2$ the  additional initial step  of local filtering  should be taken \cite{Horodeckis97} in order to achieve    a density matrix \cite{Gisin, HORO} with $F_{EPR-}>1/2$, where $\left|\psi_{EPR-}\right\rangle=1/\sqrt{2}\left(\left| 01\right\rangle-\left| 10\right\rangle\right)$. Then simple local unitary operations are applied to convert the locally filtered state to    a state with $F_{EPR}>1/2$.
In what follows we show how the knowledge about the BSA of a mixed state with $F_{EPR}<1/2$ may help  to skip the local filtering step.

\begin{figure}[h] \includegraphics[width=0.4\textwidth]{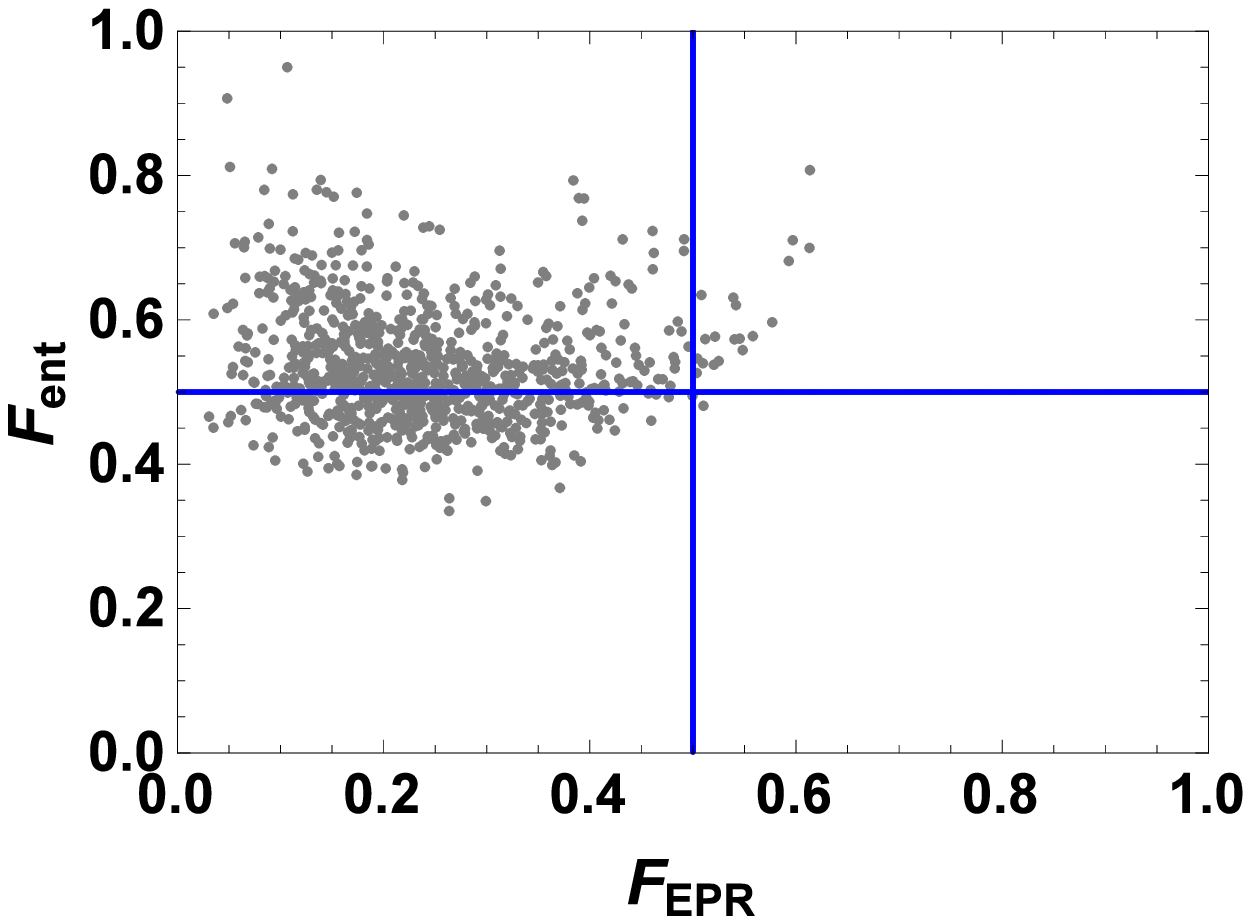}\caption{ Two-qubit NPT random states: the fidelity    $F_{ent}\left( \hat{\rho} \right)=\mathrm{Tr}\left[ \hat{\rho} \left|\psi_{ent}\right\rangle \left\langle \psi_{ent}\right| \right]$ versus $F_{EPR}=\mathrm{Tr}\left[ \hat{\rho} \left|\psi_{EPR}\right\rangle \left\langle \psi_{EPR}\right| \right]$.
We have used in total $1200$ NPT states, $600$ generated according to flat and $600$  according to induced measure.} \label{fig4}\end{figure}

As we have observed in Fig~\ref{fig3}~(b), the essentially entangled component $\left| \psi_{ent}\right\rangle$ in (\ref{for}) is a maximally entangled state for a big part ($\approx 72\%$) of tested density matrices. For the states $\widehat{\rho }$ where the latter holds true, i.e. the concurrence of $\left| \psi_{ent}\right\rangle$ is greater than $0.99$,  we calculate  the fidelity $F_{ent}$ of the random density matrices with $\left| \psi_{ent}\right\rangle$,    and we plot it versus  $F_{EPR}$ in  Fig ~\ref{fig4}. One can observe that in many cases (for the states on the upper-left  quartile), $F_{ent}>1/2$ while $F_{EPR}<1/2$. According to our statistics on $1200$ random states, no knowledge of $BSA$ requires the application of local-filtering protocol in $\approx 99\%$ of cases, while a knowledge of the BSA states requires the application of local-filtering protocol only in $\approx 50\%$ of cases. For the rest of $50\%$ it is sufficient to apply local unitary operations converting $\left| \psi_{ent}\right\rangle$ to  $\left|\psi_{EPR}\right\rangle$.

\end{document}